\documentclass[sigconf]{acmart}
\setcopyright{acmcopyright}
\copyrightyear{2024}
\acmYear{2024}
\acmDOI{XXXXXXX.XXXXXXX}

\usepackage{tikz}
\usetikzlibrary{shapes, positioning, arrows, fit, calc, backgrounds}

\acmConference[Communications of the ACM]{CACM}{2024}{}
\acmPrice{15.00}
\acmISBN{978-1-4503-XXXX-X/18/06}

\usepackage{graphicx}
\usepackage{amsfonts}
\usepackage{url}
\usepackage{graphicx}
\usepackage{color, colortbl}	
\usepackage{centernot}
\usepackage{algorithmicx}
\usepackage[linesnumbered, ruled, vlined]{algorithm2e}
\usepackage{multirow}

\newcommand{\pleq}{\preccurlyeq}

\newcommand{\vf}[1]{{\em #1}}

\newcommand{\hindex}{{\em h}-index}
\newcommand{\antichain}[1]{$#1$-antichain}
\newcommand{\lantichain}{\antichain{l}}
\newcommand{\antichains}[1]{$#1$-antichains}

\newcommand{\eat}[1]{}

\newcommand{\card}[1]{\mid\!#1\!\mid}

\definecolor{dennis}{RGB}{184, 111, 60}

\definecolor{LightCyan}{rgb}{0.88,1,1}
\definecolor{PaleGreen}{rgb}{0.59,0.98,0.6}
\definecolor{Rose}{RGB}{255,240,245}
\definecolor{PaleYellow}{RGB}{249,206,108}

\newcommand{\ccc}{\cellcolor{LightCyan}}
\newcommand{\ccr}{\cellcolor{Rose}}
\newcommand{\ccg}{\cellcolor{PaleGreen}}

\begin{document}

\title{Can we measure the impact of a database?}

\author{Peter Buneman}
\affiliation{%
  \institution{University of Edinburgh}
  \city{Edinburgh}
  \country{UK}
}
\email{opb@ed.ac.uk}

\author{Dennis Dosso}
\affiliation{%
  \institution{Seco Mind}
  \city{Padua}
  \country{Italy}}
\email{dennis.dosso@secomind.com}

\author{Matteo Lissandrini}
\affiliation{%
  \institution{University of Verona}
  \city{Verona}
  \country{Italy}
}
\email{matteo.lissandrini@univr.it}

\author{Gianmaria Silvello}
\affiliation{%
  \institution{University of Padua}
  \city{Padua}
  \country{Italy}}
\email{gianmaria.silvello@unipd.it}

\author{He Sun}
\affiliation{%
  \institution{University of Edinburgh}
  \city{Edinburgh}
  \country{UK}}
\email{h.sun@ed.ac.uk}

\begin{abstract}
In disseminating scientific and statistical data, on-line databases have almost completely replaced traditional paper-based media such as journals and reference works. 
Given this, can we measure the impact of a database in the same way that we measure an author's or journal's impact?  
To do this, we need somehow to represent a database as a set of publications, and databases typically allow a large number of possible decompositions into parts, any of which could be treated as a publication.

We show that the definition of the {\hindex} naturally extends to hierarchies, so that if a database admits some kind of hierarchical interpretation we can use this as one measure of the importance of a database; moreover, this can be computed as efficiently as one can compute the normal {\hindex}.
This also gives us a decomposition of the database that might be used for other purposes such as giving credit to the curators or contributors to the database.  
We illustrate the process by analyzing three widely used databases.


\end{abstract}




\maketitle

\section{Introduction}
\label{sec:intro}
It is almost self-evident that databases exist to publish data,  and this is undoubtedly the case for scientific and statistical databases, which have largely replaced traditional reference works. 
Database and web technology has led to an explosion in the number of databases 
that support scientific research for obvious reasons: databases provide faster communication of knowledge, they hold larger volumes of data, they are more easily searched, and they are both human and machine-readable. 
Moreover, they can be developed rapidly and collaboratively by a mixture of researchers and curators.  
For example, more than 1500 curated databases are relevant to molecular biology alone~\cite{Imker2018}.
The value of these databases lies not only in the data they present but also in how they organize that data. 

{\em If we want to measure the impact of a database, can we use its organization to treat it in the same way that we treat any other publishing agent, such as a journal or an author? }

In the case of an author or journal, most bibliometric measures are obtained from the citations to the associated set of publications. 
For a database, there are typically many ways of decomposing it into publications, so we might use its organization to guide in the choice of decompositions.
We shall show that, when the database has a hierarchical structure, there is a natural extension of the h-index that works on hierarchies.

Although the main results presented in this paper are the evaluations of the {\hindex}es of some well-known databases, this was not the original motivation. 
One of the authors was involved in a project~\cite{buneman2019summ} to automatically generate a set of conventional papers to give credit -- as authors -- to the 1000 or so researchers who had contributed to a database~\cite{harding2018iuphar}. 
By creating one publication for the whole database -- as might happen with {\em data papers} which periodically publish data summaries and are citation proxies for databases~\citep{CandelaEtAl2015}, we generate a document with an unhelpfully huge number of authors. Also, these authors receive only one additional publication credit, regardless of whether they contributed to numerous sections of the database or just a single part. 
On the other hand, generating thousands of documents, one for each "object" in the database, is almost equally useless. 
While the authors are now associated with their areas of expertise, they are unwittingly guilty of having minimal publishing units. 
The tension between those two extremes underlies the rationale for the  {\hindex}~\citep{hirsch2005index}, so the obvious question is can we extend the {\hindex} to work on databases and -- given that there is a hierarchical structure -- is there a natural extension of the {\hindex} to hierarchies?
We then have at least one non-trivial measure of the importance of a database.
However, even if this measure is not of interest, the decomposition that produced it may well be of interest, perhaps as a starting point to the curators, who want to find a useful set of publications to associate with the database for the benefit of the contributors or curators who would like to have their work properly cited.

Hierarchies are used frequently in curated databases. 
As in the three examples we use in this paper~\cite{harding2018iuphar, DrugBank2018,schoch2020ncbi}, they are  based on some kind of hierarchical classification scheme, taxonomy or ontology.   
Moreover, data sets based on a file system or data format such as JSON or XML have an intrinsic hierarchical structure (we shall refer to all of these as databases).  
We also note that, when two very similar papers are published (e.g., a preprint and its final peer-reviewed version), citation analyzers such as Google Scholar have an option to transfer citations from one paper to the other (the parent).  
Given that a paper can have at most one parent and that the parent relationship is acyclic, there is at least a partial hierarchy (a set of tree like structures) already present in the data structure maintained by the software. 

\begin{figure}[t]
  \includegraphics[width=1\columnwidth]{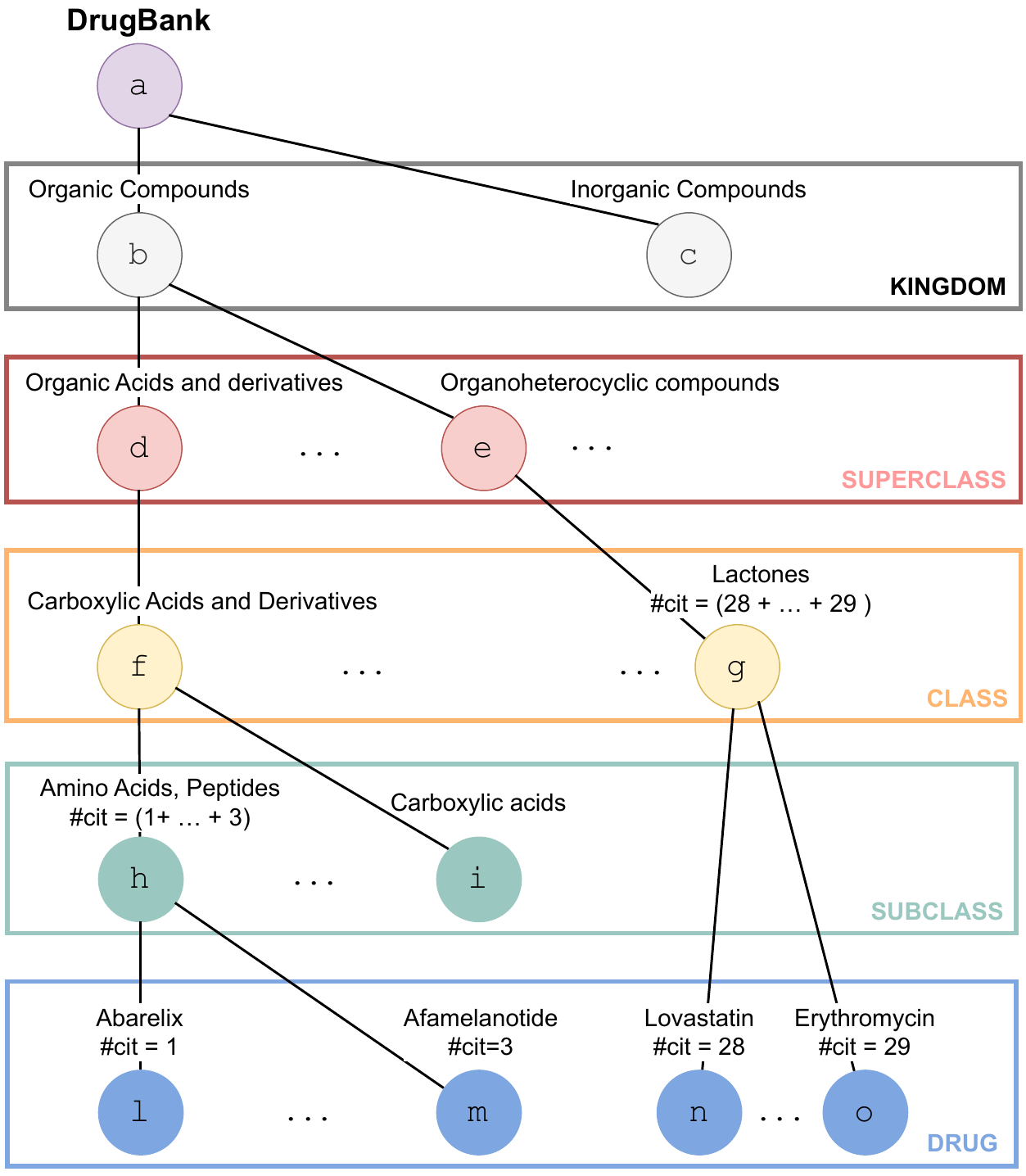}
  \caption{Partial representation of the Drugbank hierarchical structure. The web pages associated with the nodes are shown on the left. In Drugbank only the leaves of the hierarchy are directly cited; e.g., we can see that the citations of "Lactones" is the aggregation of the citations to the drugs (we report only Lovastatin and Erythromycin) belonging to the class.}
  \label{fig:drugbank}
\end{figure}

Given a hierarchy, how do we use it to limit the possible decompositions into sets of publications? 
Consider hierarchy in the  DrugBank database shown in Figure~\ref{fig:drugbank}.  
In this database, citations are only to the terminal nodes or leaves of the tree.  
If we want a citation count for some higher-level node, we would use \emph{the sum of the citation counts of the leaves below that node}, much as we would take the citation count for an issue of a journal to be the sum of the citation counts of the papers in that issue. 
Now, we propose to use a subset of these nodes as a possible decomposition; however we cannot use an arbitrary subset, because it allows double-counting 
 of citations if we allow a node and one of its ancestors to appear in the same decomposition.  
Thus we restrict our attention to {\em antichains} of nodes: an antichain is a set of nodes in which no node is an ancestor of another node in that set.  
Looking at the Linnean-style stratification in Figure~\ref{fig:drugbank}, the kingdom, superclass and class nodes each constitute antichains as do the drug nodes (the subclass nodes can have other subclass nodes as ancestors and do not). 
Also, the root node represents the database as a whole, and the leaf nodes represent the individual drugs, which both constitute hierarchies.

Now, the  {\hindex}, which is used almost universally to measure an author's output and often the importance of a journal;  it is one of the few metrics that measure both the productivity and the citation impact of authors~\citep{Silva2018}.  
It is defined as follows:
\begin{quote}
Given a set $S$ of publications, its {\hindex} is the largest number $n$ for which there is a subset of $S$ of size $n$ in which each publication has at least $n$ citations. 
\end{quote}
 The extension to hierarchies is immediate:
 \begin{quote}
Given a {\em hierarchy} $H$ of publications, its {\hindex} is the largest number $n$ for which there is an {\em antichain} in $H$ of size $n$ in which each publication has at least $n$ citations. 
\end{quote}

We will formalize this in the following section, but we should make some important observations immediately.  
First, when the hierarchy is ``flat'' (there is no ordering), the two definitions coincide. 
Second, although a set has an exponential number of subsets, its {\hindex} can be efficiently computed by sorting the set.  
Similarly, a hierarchy can have an exponentially large number of antichains, and we will demonstrate that, by using the appropriate data structures, the {\hindex} of a hierarchy can be evaluated with the same efficiency. 
This is crucial if we are going to apply it to large databases.

Third, using a hierarchy to constrain the possible decompositions is essential. 
For example,
 based on a reduction from the Partition problem, one can show that the more general problem of finding the maximal {\hindex} under an arbitrary partition is strongly \textsf{NP}-hard~\cite{dekeijzer2013hindex}, and there are algorithms and complexity research on improving and maximizing the citation indices~(e.g., {\hindex}, $g$-index, and $i10$-index) under different merging rules~\cite{dekeijzer2013hindex,van2016h,pavlou2016manipulating}.  

 Finally, much of this research can be seen as an attempt to ``game'' the {\hindex} — finding some merging strategy that enhances one's {\hindex}. 
 If, for a hierarchical organization, we equivalently define the {\hindex} of a hierarchy as the maximal {\hindex} of any antichain within the hierarchy, then it might seem as though we are also engaged in gaming. 
 However, this is not the case; we are merely employing what appears to be the most natural extension of the {\hindex} definition to hierarchies.

In the following sections, we first formalize the problem and develop some simple results on antichains that
allow that the maximal \hindex\ to be computed efficiently.  
Following this, we give three examples of databases from which we can extract an {\hindex} and look at some of the details of their hierarchical organization.  
We conclude with some speculation on how these results might be further developed. 

There are two major caveats to what we have to offer.  
The first is that we are not going to comment on the justification or fairness of the {\hindex} and its variants~\citep{bornmann2007we, yong2014critique};
we only observe that it is almost universally adopted in measuring the impact of both authors and journals and -- as we have noted above -- it may be useful in helping to decompose a database into citable units.
\eat{A single metric is not the only way to measure quality and impact, even for traditional publications; and data citation may offer further possibilities. }

Second, the results we present here are preliminary. 
As we have already remarked, the practice of data citation is still in its infancy; people still fail to cite databases or cite them improperly. 
For comparison, we have included the results of using URLs as citations.
{\em Hence, the results we give should in no way be taken as an accurate measure of the importance of the databases we examine.
They demonstrate appropriate techniques and the feasibility of measuring the {\rm \hindex}\ at scale.}

\section{The \hindex\  of a hierarchy}
\label{sec:methods}

\subsection{Hierarchies and antichains}

\begin{figure}[]
  \includegraphics[width=1\columnwidth]{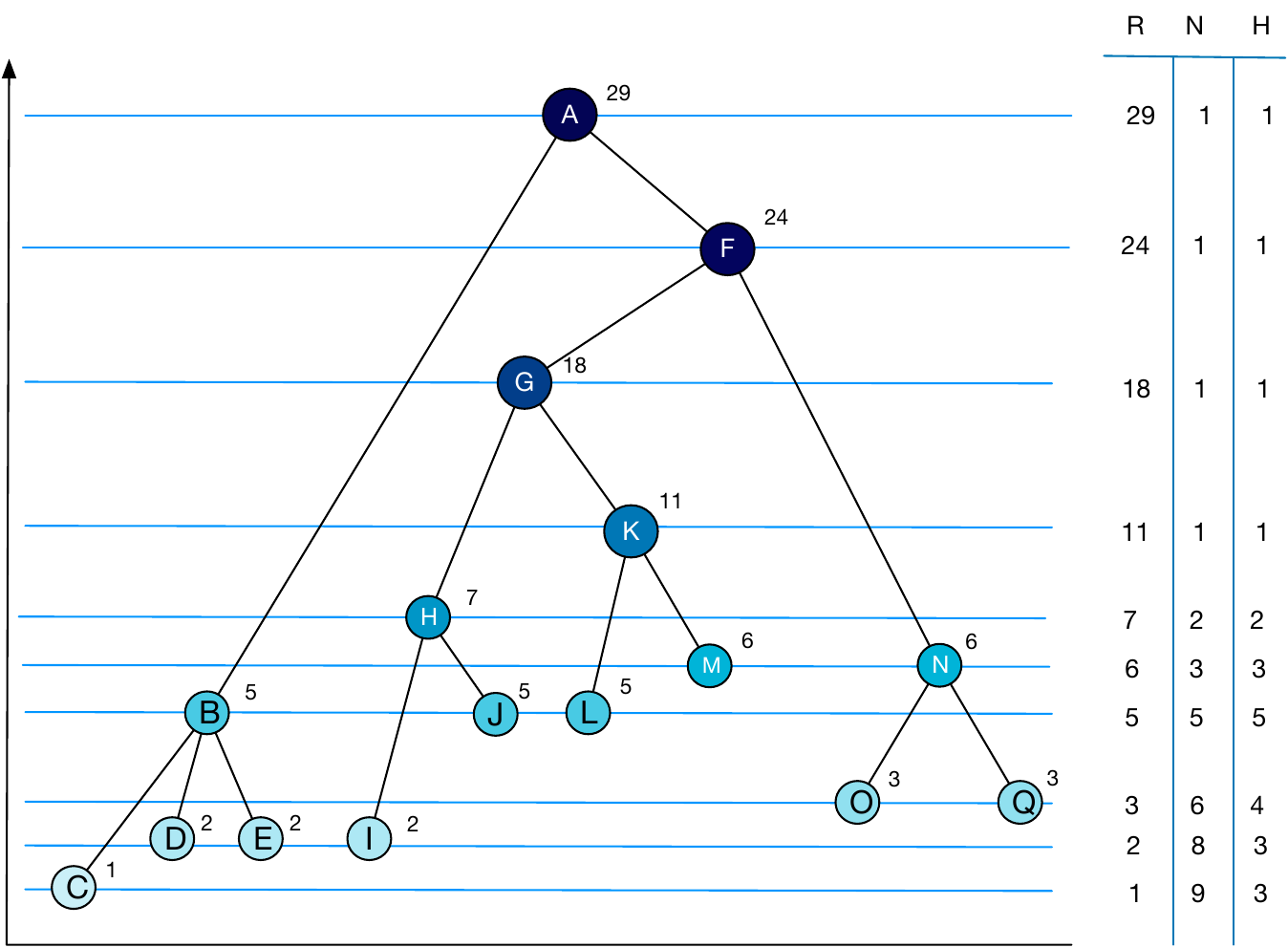}
  \caption{A hierarchical citation structure in which the level of each node is determined by its rank --  the number of citations to it. Columns: R -- the level; N -- the length of the maximal \lantichain\ at that level; and H -- the \hindex\ for that antichain.
}
  \label{fig:example_hierarchy}
\end{figure}

Our starting point is that we can model a database as a hierarchy together with a citation count for each node in the hierarchy. 
Our goal is to select a subset of nodes, aggregate the citations associated with those, and thus compute the {\hindex} of such set of nodes.
In choosing a subset of nodes, we cannot double-count citations.   
For example, in Figure~\ref{fig:drugbank} if we decide -- to compute an {\hindex} -- that the \emph{Lactones} node (node \texttt{g} at the ``classes'' level) is to be a publication, the citations to that node include all the citations to drugs beneath that node, so we cannot use a citation to one of those drugs as a citation to some other node. 
This means that we limit our candidate sets to those in which no node is an ancestor of another, or {\em antichains}.

To illustrate this,  Figure~\ref{fig:example_hierarchy} shows a hierarchy in which the root $\{A\}$ constitutes an antichain, as does $\{B, G, N\}$ and also the leaves.   
In this diagram, the level of a node is determined by its citation count or {\em rank}. 
Even though the number of antichains in a hierarchy can be exponentially large, to compute the {\hindex}, we can use the rank to cut down the number of antichains we need to examine.
For this, we need a small amount of formal development.

\begin{definition}
	A \emph{hierarchy}  is a partial order $(H, \pleq)$ 
	such that for all $a, b, c{\in}H$, $a{\pleq}b$ and $a{\pleq}c$ implies that $b{\pleq}c$ or $c{\pleq}b$. 

\end{definition}
This definition of hierarchy ensures that a node can have at most one parent. We note that $n'$ is the \emph{parent} of a node $n$ if it is the smallest node $n' \in H$, distinct from $n$, such that $n \pleq n'$.



\begin{definition}
A set $A \subseteq H$ is called \emph{antichain} if, for any distinct pair $n_1, n_2 \in A$, neither $n_1 \pleq n_2$ nor $n_2 \pleq n_1$.
\end{definition}

\begin{definition}
A {\em ranked hierarchy} $(H, \pleq, r)$ is a hierarchy $(H, \pleq)$
together with a rank (or level) function $r: H \rightarrow \mathbb{N}$, such that for all $n_1, n_2 \in H$, if $n_1 \pleq n_2$ then $r(n_1) \leq r(n_2)$.

\end{definition}

Given a subset $S$ of $H$, we will use the term \emph{minimal elements} of $S$ to refer to those nodes $n \in S$ for which there is no $n' \in S$, distinct from $n$, such that $n' \pleq n$. That is, the minimal elements of $S$ are the lowest nodes in the hierarchy belonging to $S$.
 \begin{definition}
	Given a ranked hierarchy $(H, \pleq, r)$, an \emph{\antichain{l}} in H is an antichain $A$ such that for all $n{\in}A$:
\begin{enumerate}
	\item $r(n) \geq l$,
	\item $\forall~ n'{\in}H$ if $n'{\prec}n$, then $r(n'){<}l$.
\end{enumerate}
\end{definition}

In other words, an \lantichain\ is a set in which each node has rank at least $l$  and all its children have rank less than $l$.   In Figure \ref{fig:example_hierarchy} the set $\{H, K\}$ is a \antichain{7} and the leaves form a \antichain{1}. Note that a leaf (a node with no children) with rank at least $l$ could be a member of a \antichain{l}.

We shall use the term \emph{rank-minimal} to describe the elements of a set that are minimal with respect to the rank function $r$, i.e., those elements $n \in S$ for which $r(n)$ is minimum for $S$.  

The following result guarantees  that, given an antichain with minimal rank $l$, it is possible to find an \lantichain\ which has no fewer elements.

\begin{proposition}\label{prop:rankminimal}
\label{lemma:1}
	Let $A$ be an antichain in a ranked hierarchy $(H,\pleq,r)$ with a rank-minimal node of rank $l$.
	Then, there exists an \lantichain\ $A' \subseteq H$ such that $|A'| \geq |A|$.
\end{proposition}  

The proof is straightforward and given in the additional material~(\ref{propa:2}); one constructs the antichain by a process that is similar to the top-down algorithm, which we describe shortly.

From this, we can obtain the main result.   The \hindex\ of an antichain $A\subseteq H$ in a ranked hierarchy $(H, \pleq,r)$ is defined as $h(A) = \max\{\card{S}\  \mid\  S\subseteq A \wedge \forall n\in S. r(n) \geq \card{S}\}$.

\begin{proposition}
\label{prop:1}
For any antichain $A$ in a ranked hierarchy, there is an \antichain{l}\ $A'$ such that $h(A') \geq h(A)$.
\end{proposition}

\begin{proof}  Let $l = h(A)$ be the \hindex\ of $A$. Then there is a subset $S\subset A$  for which $r(A) \geq \card{S} = l $ for all $n\in S$. A rank-minimal element of S will have rank $l'\geq l$ and Prop.~\ref{prop:1} gives us an \antichain{l'} $S'$ with at least as many elements as $S$, i.e., $S'$ has an \hindex\ no less than $l$.\end{proof}

The importance of this result is that we only have to search maximal \antichains{l} to find the  \hindex\ of any antichain in $H$.   The number of such antichains is not greater than the number of nodes, which guarantees us at least a polynomial time algorithm.  In fact, by pursuing a top-down approach, we can obtain a very efficient algorithm which, once the hierarchy is constructed, will, in practice,  only visit a subset of the nodes in the hierarchy.

\subsection{A top-down algorithm}

Consider the \antichains{l} in Figure~\ref{fig:example_hierarchy}. At the root, we have a \antichain{29} ${A}$ of length 1 and below it a \antichain{24} ${F}$ of length 1. Proceeding downwards, the first longer antichain is the \antichain{7} $\{H,K\}$ of length 2. Further down we find the \antichain{5} $\{B,J,L,M, N\}$ of length 5.  This is the \antichain{l} of greatest \hindex: it has 5 nodes, all with rank greater or equal to 5.  As we proceed downwards, the level is strictly decreasing while the length of the antichains is non-decreasing.  When the latter overtakes the former, we have just "overshot" the antichain that yields the maximal \hindex, and we can stop. Note that we have stopped before examining all the nodes in the hierarchy.

In the following we use the notation $MIN_{\pleq}(S)$ for the minimal elements of a subset $S \in H$, i.e., an antichain.
$MAX_{\pleq}$ is defined similarly. 
An \lantichain\ can thus be defined as $MIN_{\pleq}\{n \in H | r(n) \geq l \}$. 

In order to compute successively lower antichains, the algorithm maintains, for decreasing values of $l$, the following two antichains:
\begin{enumerate}
	\item $LCHAIN = \min_{\pleq}\{n \in H | r(n) \geq l\}$, i.e., the \lantichain\ of maximum cardinality among all $l$-antichains;
	\item  $DCHAIN = \max_{\pleq}\{n \in H | r(n) < l\}$, i.e., the antichain of nodes that are maximal (with respect to $\pleq$) in the set of nodes with rank less than $l$.
\end{enumerate}  

These two antichains have the following properties:
\begin{enumerate}
	\item the children of any element in $LCHAIN$ are contained in $DCHAIN$;
	\item the next lowest rank $l' < l$ of any node in the hierarchy is possessed by at least one node in $DCHAIN$;
\end{enumerate}

Also, we observe that: i) a node $n$ is in $LCHAIN$ iff $r(n){\geq}l$ and for all children $m$ of $n$, $r(m){<}l$; ii) a node $n$ is in $DCHAIN$ iff $r(n){<}l$ and the parent $p$ of $n$ is such that $r(p){\geq}l$.

\begin{center}
	\begin{algorithm}[]
		 \DontPrintSemicolon
		 \SetKwInOut{Input}{Input}\SetKwInOut{Output}{Output}
		 \Input{roots: a list of the roots of the hierarchy;\\
		 children: a function that gives the list of children of a node;}
		 \Output{An antichain with maximal \hindex\ }
		 LCHAIN $\leftarrow \emptyset$\;
		 DCHAIN $\leftarrow $ roots\;
		 $l \leftarrow max\{ r(n) | n \in DCHAIN \}$ \;
		 \While{$|$LCHAIN$| \leq l$}{
		 	oldLCHAIN $\leftarrow$ LCHAIN.copy() \;
		 	\While{$\exists n \in $ DCHAIN such that $r(n) = l$}{
		 		DCHAIN $\leftarrow $ DCHAIN $ -~ \{n\}$\;
		 		LCHAIN $\leftarrow$ LCHAIN $\cup~ \{ n \}$\;
		 		\ForEach{$c \in children(n)$}
		 		{
		 			DCHAIN $\leftarrow$ DCHAIN $\cup~ \{ c \}$\; 
		 		}
		 	}
		 	\While{$\exists n \in $ LCHAIN such that $\exists c \in $ children($n$) such that $r(c) \geq l$ }{
		 		LCHAIN $\leftarrow$ LCHAIN $-~ \{n\}$
		 	}
		 	$l \leftarrow max\{ r(n) | n \in \mbox{DCHAIN} \}$ \;
		 }
		 \textbf{return} oldLCHAIN\;
	\caption{Finding $l$-minimal antichains}
    \label{algorithm}  
	\end{algorithm}
\end{center}

Algorithm \ref{algorithm} details the procedure used to find the antichain that guarantees the maximum \hindex. On lines 3 and 13, the nodes with the highest rank (i.e., citation count) $l$ are found in the $DCHAIN$. These nodes are removed from the DCHAIN (line 7) and added to the $LCHAIN$ (line 8). The children of these nodes are then added to $DCHAIN$ (lines 9 and 10).
Then, in lines 11 and 12, all the non-minimal nodes in $LCHAIN$ are removed, making it a proper $l$-antichain. 
The process continues until the cardinality of $LCHAIN$ is bigger than the highest rank $l$ in $DCHAIN$. At this point, each node in $DCHAIN$ has a rank $\leq l$. Thus even if it were to be moved from $DCHAIN$ to $LCHAIN$ there would not be an increase in the value of the \hindex\ for $LCHAIN$. At this point, the algorithm can stop.

\subsection{Algorithm discussion}
\label{sec:performance}
Referring to Algorithm \ref{algorithm}, the most obvious thing to do is to use heaps to implement $DCHAIN$ and $LCHAIN$.
The heap values for each node \vf{DCHAIN} should be the node rank, while the values for those in  \vf{LCHAIN} should be the maximal rank of the children of the node, which can be computed at line 9.
With these representations, the ``search'' operations on lines 3, 6, 11, and 13 all incur unit cost, while the insert or delete
operations on lines 7, 8, 10, and 12 are all $O(\log n)$ in the length of the antichain. This gives a bound of $O(n\log n)$ in the hierarchy size for the whole algorithm.

A further observation is that at line 9, we need only consider child nodes for inclusion if their rank is greater
than the current length of \vf{LCHAIN}; if this is not the case, they cannot participate in the final antichain, and they
together with their descendants in the hierarchy may be safely ignored.  
In describing the results for the various databases in Section~\ref{sec:results} we have included two figures: visited (vis.) -- the number of nodes whose rank was interrogated and digested (dig.) -- the number of nodes that entered \vf{DCHAIN}. 
The algorithm's running time will be bounded by a constant in vis. and a constant times $n\log n$ in dig. 
In the measurements for a whole hierarchy (not the truncated ones where we cut off the hierarchy at a certain level), the digested nodes accounted for less than $10\%$ of the total.

Another observation is that the algorithm requires that the set of nodes as input is a hierarchy, and thus that this hierarchy is built before this algorithm is run.
Finally, we note that in the degenerate case of a hierarchy made of only leaves, the algorithm requires $O(n \log n)$ time to create the \vf{LCHAIN} heap.


%

\section{Experimental Analysis}
\label{sec:results}

Using this algorithm, we compute and analyze the computation of the {\hindex} of three widely-used databases: Drugbank which we describe in Section~\ref{sec:intro}; the  IUPHAR/BPS Guide to Pharmacology~\cite{harding2018iuphar} (GtoPdb), which is an open-access database on the biological targets of drugs and other molecules; and, the taxonomic database curated by the National Center for Biotechnology Information (NCBI), which maintains one of the most comprehensive and best-organized collections of medical data~\cite{Mcentyre2002ncbi}. 
The online additional material details all the information about these databases and how we collected the corresponding citation counts. 

\subsection{Drugbank}

\begin{table}[t]
\centering
\caption{H-index and statistics on Drugbank. "Full" (top row) -- the calculations on the full hierarchy; "Leaves/drugs" (second row) -- those on the flat hierarchy (only leaves).}
 \begin{tabular}{l||r|r|r|r||rr}

                hierarchy   & \textbf{\hindex} &  median & max & nodes   &  vis. & dig. \\
  \hline
  \ccg Full   & \ccg 69      &   \ccg 100   &  \ccg 260   &   \ccg 11,803 &  \ccg  3,743  &   \ccg 417 \\
  \ccc Leaves/drugs     & \ccc 33      &   \ccc 44   &  \ccc  76   &  \ccc 10,303 & \ccc  --   & \ccc -- \\ \hline
  Subclass         & 63      &   105   &  1,369   &      837&     541   &    276 \\
  Class            & 49      &   120   &  2,327   &     328  &    287   &    154 \\

\end{tabular}	
\label{table:drugbank_statistics}
\end{table}

The results of the analysis on Drugbank are shown in Table \ref{table:drugbank_statistics}, where \emph{median} and \emph{maximum} refer to the citation counts of the antichain that yields the given \hindex\ (the median is sometimes given in reporting the \hindex\ of a journal).  
The visited (vis.) and digested (dig.) columns refer to the discussion about Algorithm 1.
Only a few nodes are leaves in the antichain for the entire hierarchy. 
The \emph{Leaves/drugs} row shows the \hindex\ for the ``flat'' hierarchy of the drugs themselves, that is, the \hindex\ obtained considering only the drugs as a set of ``publications'' that receive citations, without considering the hierarchy of the database.  
The \emph{Class} and \emph{Subclass} rows show the results for the hierarchies obtained by removing all nodes below those strata. The table clearly shows the advantage to computing the \hindex\ on a full hierarchy and second that constraining the antichains to a given stratum reduces the \hindex. We also see that the algorithm calculates the \hindex\ visiting approximately one-third of the nodes in the entire hierarchy.

\subsection{GtoPdb}
\begin{figure}[t]
  \includegraphics[width=1\columnwidth]{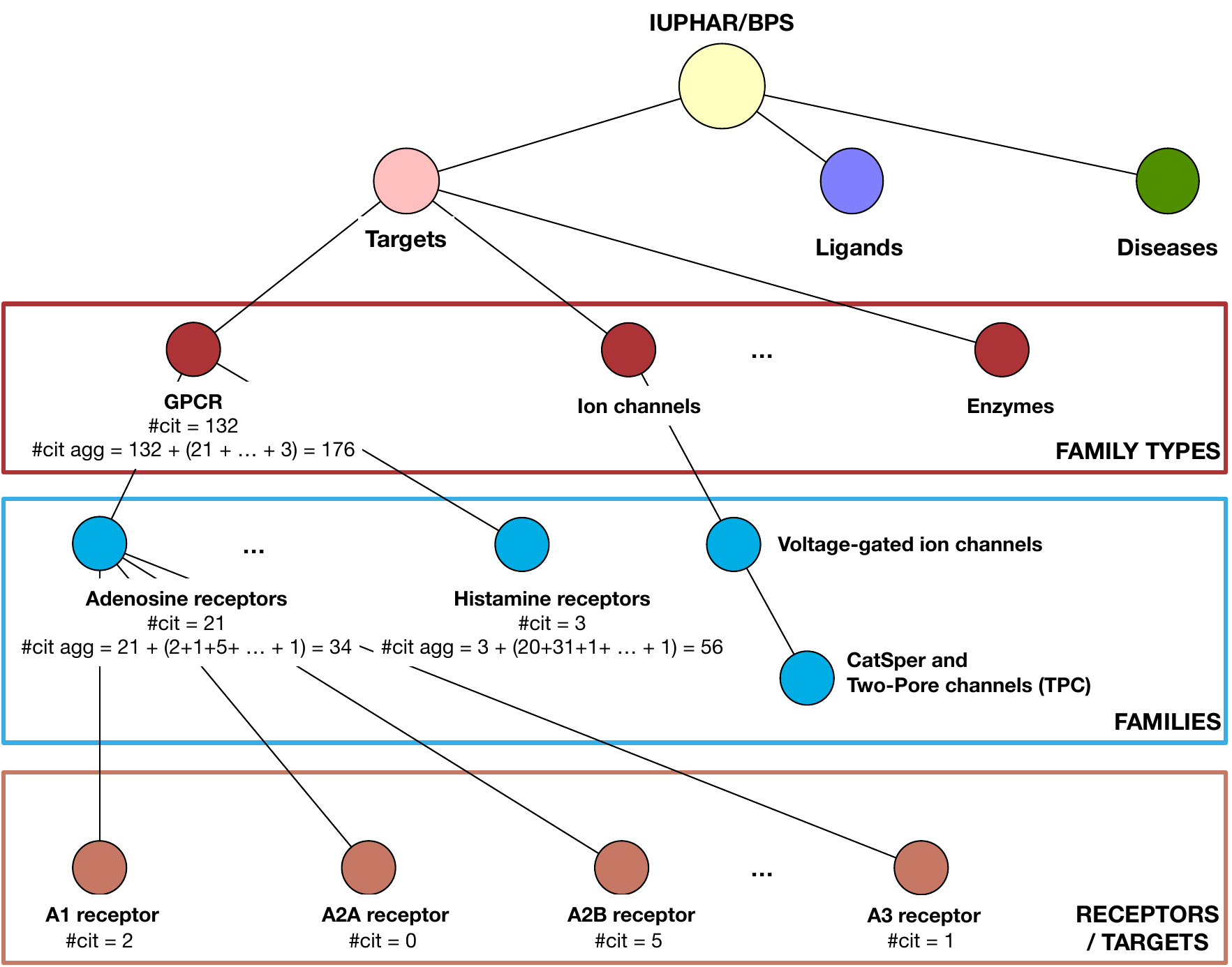}
  \caption{Part of the IUPHAR/BPS hierarchical structure. The root node represents the whole databases, and the family structure is not stratified as families may have subfamilies. All the nodes in the hierarchy can be independently cited; e.g., we show some sample citation numbers for the GPCR branch of the tree, where the internal nodes can receive direct citations (\texttt{\#cit}) that could be also aggregated (\texttt{\#cit agg}) with the citations of the child nodes.}
  \label{fig:iuphar}
\end{figure}

The  IUPHAR/BPS Guide to Pharmacology (GtoPdb) is an open-access relational database with an accompanying website on the biological targets of drugs and other molecules. As shown in Figure \ref{fig:iuphar}, GtoPdb has a hierarchical structure. It combines the expertise of approximately 1,000 researchers worldwide and was an early stimulus for data citation~\cite{buneman2010rule} to give credit to the contributors. Here, however, we are using it to assess credit to the database as a whole.

A critical difference between Drugbank and GtoPdb is that in GtoPdb, there are citations to the intermediate nodes in the hierarchy.  In fact the data for these "family" nodes often resembles short conventional papers with descriptive material and statistics on the targets (leaf nodes) included in the family.  This raises the possibility that we might want to treat a family node as independent of the nodes it describes, and we introduce a {\em lifting} transformation that allows us to do this.

\begin{table*}[t]

\centering
\caption{\hindex\ calculations for GtoPdb. "Full" (top rows) are the calculations on the full hierarchy; "No" (second rows) are those on the flat hierarchy (only leaves); and "Lifted" (bottom rows) are those on the lifted hierarchy.  }
\begin{tabular}{l|l||r|r|r|r||rr}

      Cit. type & Hierarchy & \textbf{\hindex}  &  median & max & nodes   &  vis. & dig. \\
  \hline
\multirow{4}{*}{Citations}  & \ccg Full     &    \ccg  55 & \ccg  77      & \ccg 276      & \ccg 4,079    & \ccg 1,185    & \ccg 275  \\
           & \ccc No      &  \ccc    32 & \ccc  44      &   \ccc    94 & \ccc 4,079    & \ccc  --      & \ccc --  \\ \cline{2-8}
           & Families     &      54 &   86    &   276    &   807   &   572    & 253 \\
           & \ccr Lifted  &   \ccr   55 &  \ccr 77      &\ccr 276      & \ccr 4,872    &  \ccr 1,185    & \ccr 275  \\
\hline\hline
\multirow{4}{*}{Links}      & \ccg Full     &   \ccg  167 & \ccg 251      & \ccg 880      & \ccg 4,079    & \ccg 1,584    & \ccg 419  \\
           & \ccc No   \ccc   &172    \ccc   & \ccc 253   & \ccc 909       & \ccc 4,079   & \ccc --       & \ccc --   \\ \cline{2-8}
           & Families     &  135    &  298     & 3,013      &   807   &  588     & 211  \\
           & \ccr Lifted   & \ccr 179    &  \ccr 265     & \ccr 909      &  \ccr 4,872  & \ccr 1,604     & \ccr 468 \\

\end{tabular}
\label{table:iupharcalcs}
\end{table*}
Table \ref{table:iupharcalcs} presents the results for the GtoPdb database divided into the horizontal sections  ``Citations''
and ``Links''. 
The first section corresponds to results obtained with citations extracted from the citing papers, while the second part is obtained by accounting for web links as citations.

 As with Drugbank, the first two rows of each section in the table show the analysis
 on the full hierarchy and on the ``flat'' collection of nodes, which
 now contains a substantial number of cited intermediate (family)
 nodes together with the leaf target nodes. The ``Families only'' shows that little is lost by removing
 the leaves from the hierarchy, thus suggesting that many citations are also given to the intermediate nodes in the hierarchy. 
 The meaning of the ``Lifted hierarchy'' row is discussed below.

 It is also interesting to note that families of nearly all heights -- where
 by height, we mean the maximum number of edges to a leaf node -- figure
 in the antichain that provides the \hindex\ for the full hierarchy.
 The figures are: 5 at height 3, 9 at height 2, 4 
 to leaves, and the remaining 37 to bottom level families at height 1.

\begin{figure}[]
  \begin{center}
    \includegraphics[width=1\columnwidth]{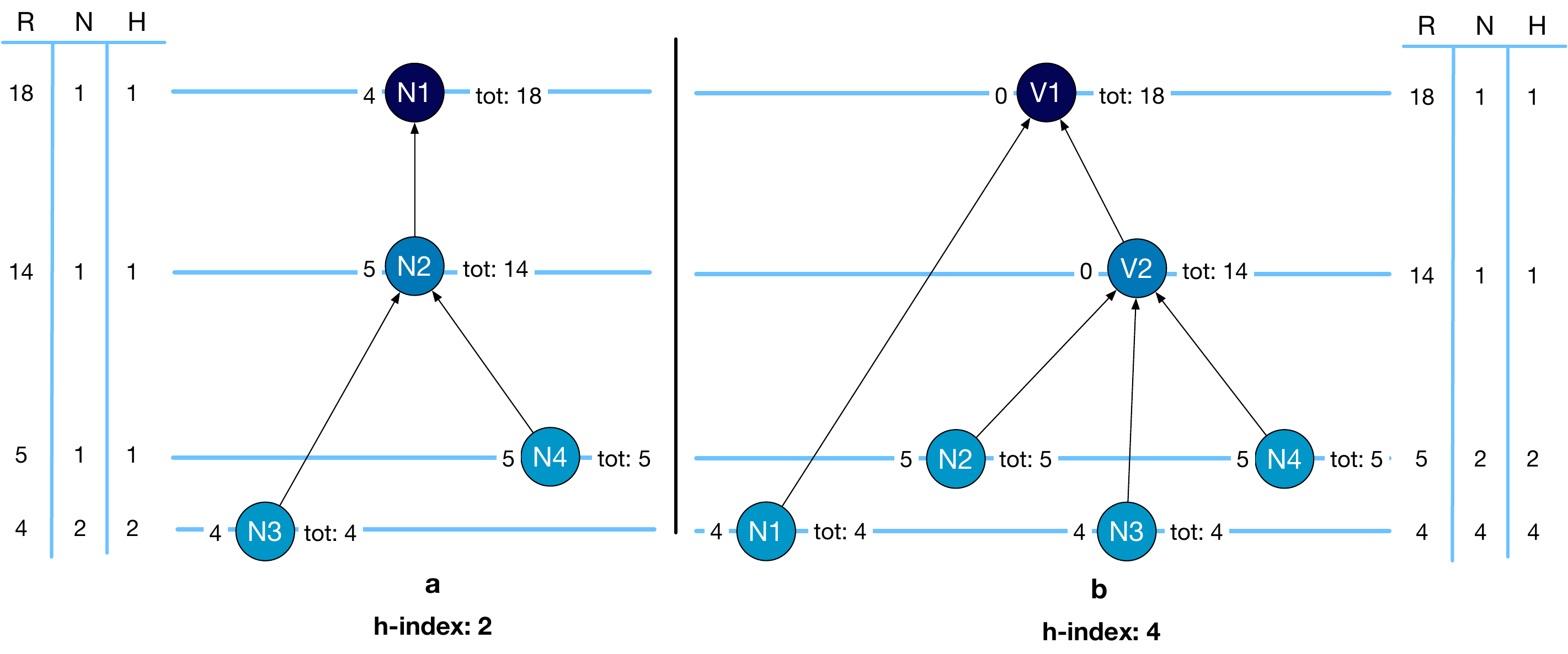}
  \end{center}
  \caption{Example of one hierarchy (a) and its corresponding ``lifted'' version (b). Once again, R is the value of the rank at each level, N is the cardinality of the maximum antichain at each level, H is the \hindex\ of that antichain. Below each hierarchy, its \hindex\ is highlighted.}
  \label{fig:lifting}
\end{figure}

\medskip
{\noindent}\textbf{Lifting.}
For reasons mentioned in the additional material, the Links section of Table \ref{table:iupharcalcs}, is based on highly skewed citation counts, and the results are almost meaningless.   We have included it in this section because it shows an interesting phenomenon.
We see -- perhaps surprisingly -- that computing the \hindex\ on the full hierarchy gives a {\em lower} figure than that for the set of nodes with no structure (167 vs 172).
It appears that creating a hierarchy on a set of publications, in this case the GtoPdb web pages, and summing the citation counts
as we have proposed, does not necessarily increase the \hindex\ with respect to the simple set of all the publications.  

Consider now the hierarchy in Figure~\ref{fig:lifting}.a. Each node is annotated on the left with its number of ``direct'' citations and on the right with the sum of the number of citations to that node and to all of its descendants.
It is easy to see that, for this first hierarchy, the longest antichain is the set $\{N_3, N_4\}$, which has an
\hindex\ of 2. 
However, had we considered the set of nodes $\{N_1, N_2, N_3, N_4\}$ without a hierarchy and with a number of citations as the one obtained by direct citations to them ($4, 5, 4, 5$ respectively), its \hindex\ would be 4. 
Therefore, in this example, the use of the hierarchy does not bring the highest \hindex\ that could be obtained with the available publications. 

On the other hand, had Figure~\ref{fig:lifting}.a been part of a larger hierarchy, it might have been advantageous to have a node such as $N_1$ with a relatively high rank. 
The implication is that, in a bigger hierarchy, new and longer antichains could have been found, where $N_1$ could contribute to the generation of a greater \hindex.
The problem is that in the computation of the \hindex, as described in Section~\ref{sec:methods}, the inclusion of a node in an antichain precludes the inclusion of any ancestor or descendant node in the same antichain.  
Thus, a descendant and ancestor node cannot be considered \emph{independently} in the computation of the \hindex. 

In a case such as GtoPdb, however, the information associated with a family may complement the information
associated with an object (child) in that same family. Therefore, It is reasonable to ask that both parent and child could be counted in the computation of the \hindex{}. 

To deal with this problem, we introduce a transformation called {\em lifting}: for \emph{each} internal node $N_i$ 
in the hierarchy, a surrogate node $V_i$ is created.
$V_i$ becomes the parent of $N_i$ along with all the children of $N_i$, and $N_i$ becomes a leaf.
$N_i$'s rank is now determined only by its direct citations, and the new node $V_i$ enters with a number of direct citations equal to $0$ and a rank equal to the sum of its children's citations.

Figure~\ref{fig:lifting}.b shows the result of applying this transformation to Figure~\ref{fig:lifting}.a.
Note how the set of all non-surrogate nodes $\{N_1, N_2, N_3, N_4\}$ now can be an antichain at rank 4.

After applying this transformation in the GtoPdb hierarchy, the results shown in Table \ref{table:iupharcalcs}(``Lifted Hierarchy'' row) show an \hindex\ of 55 with data citations, and 179  with links. 

We mentioned in the introduction that, for conventional papers, when two papers are similar, one may designate one of the papers as the parent, which will then receive all the citations for the child. 
Whether it is advantageous to combine the two depends on the authors.  
The advisor with a high existing h-index might want to do this, but the student with few publications might not.  
A lifted representation of the publication hierarchy and computing the {\hindex} of an author on that hierarchy would keep both advisor and student happy!


\subsection{NCBI Taxonomy}

\begin{table*}[t]
\caption{\hindex\ calculations for NCBI Taxonomy. "Given" (first rows) refers to the NCBI taxonomy without modifications; and "Lifted" (bottom rows) are those on the lifted hierarchy. "Pubmed" (top rows) are the calculations counting only the citations from Pubmed; and, "All" are the calculations counting Pubmed and incoming web links as citations.} 
\normalsize
\begin{tabular}{c||l|l|r|r||r r}
Hierarchy& Refs from & Subtree     &  \textbf{\hindex} & Nodes & vis & dig\\ \hline
        
         & \ccc Pubmed    & \ccc animals     & \ccc 1,181   &\ccc 1,147,717& \ccc 47,880    &\ccc    5,724\\
Given    & \ccc Pubmed    & \ccc vertebrates &   \ccc 681  &\ccc 112,271 &\ccc 17,503    &\ccc    3,133\\
        
         &  \ccr All       & \ccr animals     &  \ccr 3,794  &\ccr 1,147,717&\ccr 489,692   &\ccr    16,712\\
         & \ccr All       & \ccr vertebrates & \ccr 2,065  & \ccr 112,271 & \ccr 43,282   &  \ccr   9,254\\
\hline
         & \ccc Pubmed    & \ccc animals     & \ccc 1,301   & \ccc 1,272,867&\ccc 47,665    &  \ccc 6,560\\
Lifted   & \ccc Pubmed    & \ccc vertebrates & \ccc 758   & \ccc 131,464 & \ccc 17,700    & \ccc 3,638\\
         & \ccr All       & \ccr animals     & \ccr 3,887   &\ccr 1,272,867 &\ccr 494,769   &\ccr 18,284\\
         & \ccr All       & \ccr vertebrates &  \ccr 2,148   & \ccr 131,464 & \ccr 45,996   & \ccr 10,371\\
\end{tabular}\label{ncbicalcs}
\end{table*}

Among the databases maintained by NCBI is the taxonomy database~\cite{schoch2020ncbi}, which contains ``a list of names that are determined to be nomenclatural correct or valid (as defined according to the different codes of nomenclature), classified in an approximately phylogenetic hierarchy''\footnote{\url{https://www.ncbi.nlm.nih.gov/books/NBK53758/}}.
The hierarchy contains some 2.3 million nodes or {\em taxons}, and for each node, the web interface displays incoming links from other NCBI databases and Pubmed, which can be considered as citations from conventional literature.   The structure of NCBI is similar to that of IUPHAR/BPS as all nodes in the hierarchy are citable.
There are approximately 3.2 billion incoming links for all nodes, of which 7.8 million are from Pubmed. 
For reasons described in the additional material, we confined our attention to the vertebrate and the animal sub-hierarchies. Vertebrates are a subset of animals, and animals account for approximately half the nodes in the complete hierarchy.

The results on the NCBI taxonomy hierarchy are shown in Table~\ref{ncbicalcs}, where they are divided between the results from only the incoming links from Pubmed and the results obtained using all the available links. Unsurprisingly, the \hindex\ for animals is substantially greater than that of its sub-hierarchy of vertebrates.   Moreover, measuring an \hindex\ using incoming links gives a substantially higher result and indicates the magnitudes we might see if we measure \hindex{}es using links rather than conventional citations. Lifted hierarchies noticeably raise the results.

\section{Conclusions}
\label{sec:conclusions}
This paper demonstrates that databases can be treated similarly to authors and journals for the purpose of measuring scholarly impact.
We have shown a natural and efficient way to compute the {\hindex} of a hierarchy of citable units and to apply this to computing the {\hindex} of databases. 
Preliminary results show that, for some well-known curated databases,  the {\hindex}, when measured by citations from conventional literature, is comparable with that of journals. 
When measured by links from other databases, it can be substantially higher.
The results are preliminary because data citation is not as widely practiced as it should be. 

Throughout this paper, we have deliberately avoided any discussion of the rights and wrongs of using the {\hindex} as a measure of the impact or quality of research output, which is a matter of continuing discussion~\cite{yong2014critique,brito2021inconsistency}. 
One empirical and theoretical observation is that the {\hindex} is correlated with the square root of the total number of citations. 
In fact the ratio is in the neighborhood of 0.5.  
This holds for our results when considering citations {\em from the literature}.  
However, the results are very different when we measure -- as for the NCBI taxonomy -- an {\hindex} for incoming links, whose total is in the billions.  
Here the ratio is an order of magnitude lower, indicating a skewed distribution that requires further investigation.

There are many more open problems. 
The most obvious is what one can do without an apparent hierarchy. 
If no classification scheme is available, some form of cluster analysis can nearly always find one. 
The more likely problem is that there are several classification schemes~\cite{ashburner2000gene}, and one may want to choose or combine them. 
Also, a classification scheme may not be hierarchical because a node may have more than one parent. 
Apart from coming up against the complexity issues described in Section~\ref{sec:intro}, one has the prior problem of how the citations for a given node should be distributed among its parents to achieve a sensible ranking. 
It may be that the notion of disjunctive citations~\cite{wu2018data} could help with this. 
Finally, unlike conventional publications, databases evolve over time, yet for the purpose of citation, we want to treat them as conventional, immutable publications.  
This is a general problem with data citation and will also affect how we measure the impact of databases.

\begin{acks}
The authors would like to thank the reviewers for constructive criticisms and Simon Harding, Jonathan Bard and Aris Filos-Ratsikas for their help. 
Huawei UK partially supported Peter Buneman. 
Dennis Dosso's work was carried out while he was with the University of Padua. 
Most of the research was conducted while Matteo Lissandrini was at Aalborg University.
Gianmaria Silvello was partially supported by the HEREDITARY Project as part of the European Union's Horizon Europe research and innovation programme under grant agreement No GA 101137074. He Sun was supported by an EPSRC Early Career Fellowship (EP/T00729X/1).

\end{acks}

\newpage

\bibliographystyle{ACM-Reference-Format}
\bibliography{references.bib}

\newpage
\clearpage
\noindent{\Large\sc Additional Material}
\appendix
\section{Proofs}

\begin{proposition}
\label{propa:2}
	Let $A$ be an antichain in a ranked hierarchy $H$, and let $n \in A$ be a rank-minimal node of rank $l$.
	Then, there exists an \lantichain\ $A' \in H$ such that $|A'| \geq |A|$.
\end{proposition}
\begin{proof}
Given $S_1, S_2 \subseteq H$, we say that $S_1 \sqsubseteq S_2$ if $\forall s_1 \in S_2, ~ \exists s_2 \in S_2$ such that $s_1 \pleq s_2$, i.e., all the nodes in $S_1$ are also in $S_2$ or are descendants of a node in $S_2$. 
	It is immediate to note that $\sqsubseteq$ is a partial order and, since $H$ is a hierarchy, that if $S_1 \sqsubseteq S_2$ then $|S_1| \geq |S_2|$, since the nodes in $S_2$ cover the ones in $S_1$, they can be less than those in $S_1$.
	
	Now let $A$ be an antichain with a rank-minimal node $n$ of rank $l$.
	If no node in $A$ has a child with rank $\geq l$, then $A$ is already an \lantichain. 
	Otherwise, pick a node $n'$ in $A$ with at least one child with rank $\geq l$. Replace $n'$ with all and only its children with rank $\geq l$ to obtain a new antichain $A_1$.
	We have that $A_1 \sqsubseteq A$, $A_1 \neq A$, and $|A_1| \geq |A|$. 
	By repeating this process, we obtain a strictly decreasing sequence in $\sqsubseteq$, which must terminate in an \lantichain\ $A'$ with $|A'| \geq |A|$.
\end{proof}

\begin{figure}[h]
  \includegraphics[width=\columnwidth]{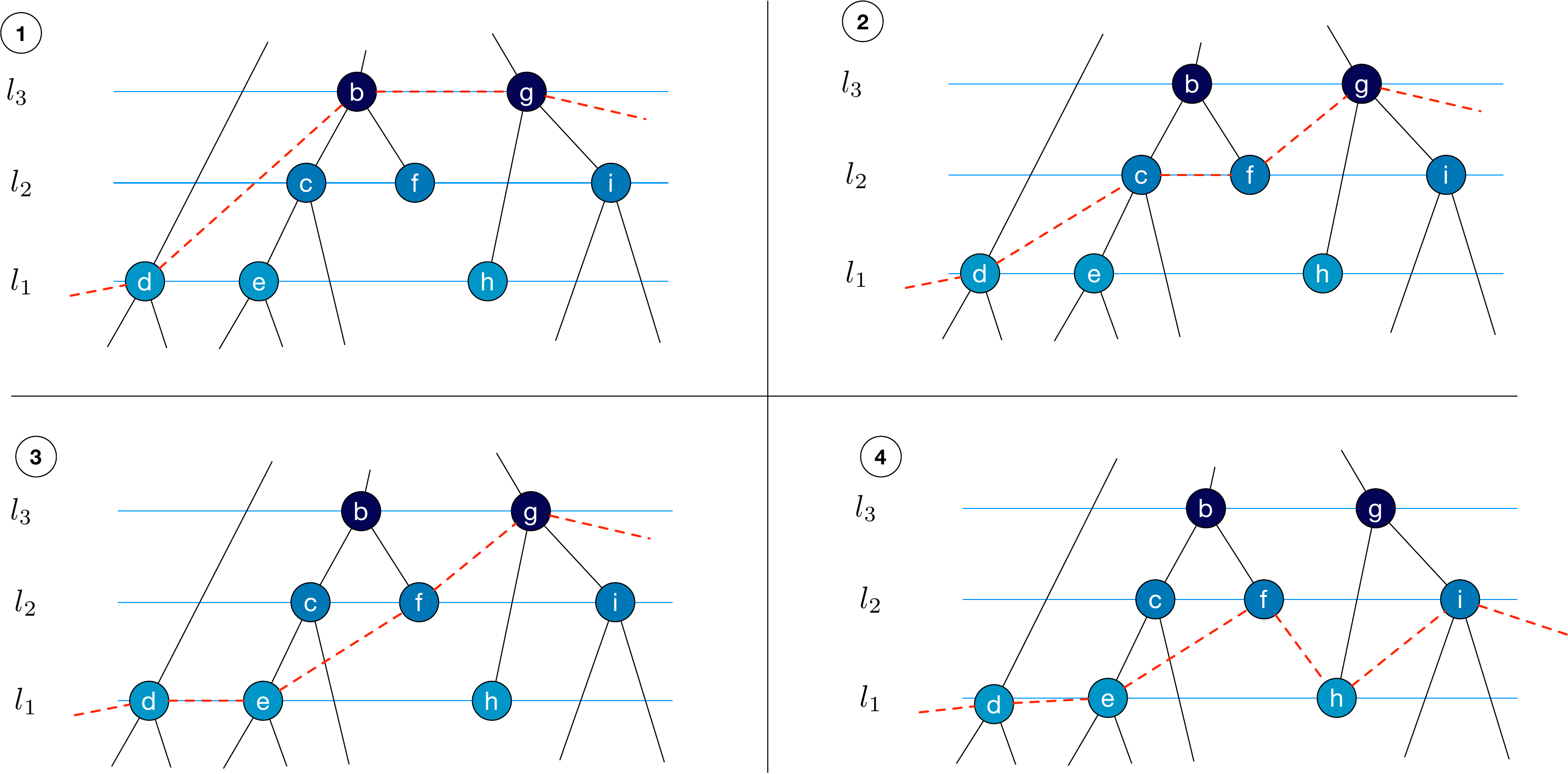}
  \caption{Example of the procedure described in Lemma \ref{lemma:1} applied on one example hierarchy.}
  \label{fig:lemma_1}
\end{figure}

As an example of the procedure used in the proof of Prop. \ref{propa:2}, refer to Figure \ref{fig:lemma_1}.
At point $1$, there is   an $l_1$-antichain $\{d, b, g\}$. Since $b$ and $g$ present children with rank $\geq l_1$, it is possible to ``expand downward'' the antichain, including their children. At step 2, $b$ is substituted with $c$ and $f$. Continuing, at step 3, $c$ is substituted with its child $e$, the only one at level $l_1$. At the last step, $g$ is substituted with $h$ and $i$. At this point there are no more children of the nodes in the $l_1$-antichain $\{d, e, f, h, i\}$ with children above the threshold $l_1$, and the process terminates. 

\section{Further details of the databases}
\label{sec:cases}


\paragraph{Drugbank}
\label{sec:drugbank}
In this commercial database of
drugs \footnote{\url{https://go.drugbank.com/releases/latest}}, the single drugs are the leaves in a taxonomy where the root is the entirety of the database, and the internal nodes are the classes of drugs. Drugbank includes a Linnean-style stratified classification scheme, as seen in the partial representation of Figure \ref{fig:drugbank}: Kingdom, Superclass, Class, and Subclass. 
In Drugbank, each drug is associated with a specific webpage with an URL like the following: \texttt{\url{https://go.drugbank.com/drugs/<id_of_the_drug>}}. 
We note that the papers referring to a Drugbank's drug often report its webpage identifier as a form of citation in the literature. Hence, using Google Scholar and the Drugbank ids of the drugs, we obtained the \emph{number of citations} to each drug webpage. As indicate in the article, the number of citations collected with this method underestimates the actual number.


\paragraph{GtoPdb}
\label{sec:iuphar}

As in  Drugbank, the leaf nodes in the GtoPdb hierarchy correspond to
individual substances.  However it differs from Drugbank in two
important respects. First, the hierarchy is not stratified into Linnean-style levels: there is a family/sub-family structure that varies in depth; currently the maximum length of a path is four.   
Second, in addition to the leaf nodes, the intermediate (family) nodes may also be cited.  For example, both the leaf nodes and the intermediate nodes may carry a citable descriptive narrative. The hierarchy, which essentially underlies the web presentation of the database, \footnote{\url{https://www.guidetopharmacology.org/}} contains $4,079$ nodes of which $3,286$ are leaves.

The database has an accompanying {\em Concise Guide}~\cite{Alexander2019concise} -- a set of publications extracted automatically from the database every two years.   
Until recently, authors using the database were encouraged to cite the concise guide when they wanted to use some data in the database. Currently, every citable unit of the database is regularly published as an article in an online journal~\cite{buneman2019summ}, which can cite and receive citations counted by citation systems like Google Scholar. 
In this work, we focus on the part of the database containing information about targets, since it is the part with an explicit hierarchical structure. In particular, by crawling the GtoPdb web pages concerning targets, it is possible to build this hierarchy following the hyperlinks in the pages. 
In GtoPdb each target (also known as receptor) is contained in a family. Each family, in turn, is contained in one of eight family types. The root of this hierarchy can be though as the part of the database dealing with targets. 

Every two years GtoPdb publishes a data paper (e.g.,  \cite{Alexander2019concise}) summing up the content of the database. While these papers describe GtoPdb as a whole, other papers focus on one of the target family types, such as \cite{Alexander2019Gprotein}. These can be considered as data papers \emph{representing} the different parts of the database, in this case the family types. 

Leveraging on this organization, we scraped Google Scholar getting the papers citing sections of above mentioned data papers about GtoPdb for the 2017-2020 timespan; hence, before the new publication/citation system was set up.

We then downloaded all (up to December 2020) the papers published in the British Pharmacological Society Journals\footnote{\url{https://bpspubs.onlinelibrary.wiley.com/}} citing one or more of the GtoPdb data papers. We then divided these papers in terms of version of the database they cite (the 2017--2018 or 2019--2020 one) and in terms of which family type they cite. This means that the same paper, citing for example both Enzymes and Kinases, is classified in both categories.
We were interested in the papers published specifically by the BPSJ since they must contain the URLs of the webpages they cited, following the editor's guidelines. 
It is therefore possible to scrape these papers looking from the citations to the webpages representing the single targets and target families, filtering these citations depending on the family type being targeted.
 This yielded the $7,870$ citations to $1,446$ of the $4,079$ nodes in the hierarchy; note that $5,611$ citations are to leaf nodes.

There is a second approach to obtaining the citations to GtoPdb, which is to count the incoming links to
the target web pages. These links may come from conventional publications or
 from other web pages or data sets. Google analytics
 counts inbound links within a given domain
and provides a list of $1,000$  of these with the highest count.
Unfortunately, many of the links in the GtoPdb domain are not to pages
of the GtoPdb database.   In fact, only about some $340$ nodes in
the database are counted with a total of approximately $75,000$ links.
This means that nodes with a low link count are not mentioned and the
results are skewed. Because of the cut-off, the least
(non-zero) cited
node has a link count of $167$.   In fact, we did not consider them in our calculations in the article.

\paragraph{NCBI taxonomy}
\label{sec:ncbitax}

The National Center for Biotechnology
Information\footnote{\url{https://www.ncbi.nlm.nih.gov/}} (NCBI) maintains what is surely the most comprehensive and best organized collection of medical data~\cite{Mcentyre2002ncbi}.  
In particular it contains a catalog, Pubmed, of much of the world's biomedical literature and a number of databases. 
Through Entrez, an integrated retrieval system, links between 2.5 billion records are maintained~\cite{sayers2019database}.  
This includes links to and from Pubmed references which could be treated as citations from conventional literature.  

In the previous two examples (Drugbank and GtoPdb) our estimates of citation counts are low both
because data citation is not widely practiced and also because it is often impossible thoroughly to ``scrape'' the literature for such citations.
With NCBI and Pubmed, we believe that we have a much more realistic count of citations and links, and thus obtained an estimation of what a real link-based \hindex\ would look like.

Unfortunately, using the API to obtain these link counts is slower than scraping them from the web interface. Therefore, we confined our attention to the subtrees of vertebrates and animals, which account for about half of the entire hierarchy.

\end{document}